\documentclass[%
reprint,
superscriptaddress,
showpacs,
 amsmath,amssymb,
pre,
floatfix,
]{revtex4-1}
\usepackage{graphicx,ifpdf}
\usepackage{bm}
\ifpdf 
\usepackage[pdfauthor={V. S. Nikolayev},pdftitle={Quench cooling under reduced gravity},
colorlinks,linktocpage,breaklinks,pdftex,hyperfigures]{hyperref} \fi

\begin{document}
\title{Quench cooling under reduced gravity}

\author{D. Chatain}
\author{C. Mariette}
\affiliation{Service des
Basses Temp\'eratures, UMR-E CEA / UJF-Grenoble 1, INAC, 17 rue des Martyrs, 38054 Grenoble Cedex 9, France}
\author{V. S. Nikolayev}\thanks{Corresponding author}
\email{vadim.nikolayev@espci.fr}
\affiliation{Service des
Basses Temp\'eratures, UMR-E CEA / UJF-Grenoble 1, INAC, 17 rue des Martyrs, 38054 Grenoble Cedex 9, France}
\affiliation{ESEME, PMMH-ESPCI, 10, rue Vauquelin, 75231 Paris Cedex 5, France}
\author{D. Beysens}
\affiliation{Service des
Basses Temp\'eratures, UMR-E CEA / UJF-Grenoble 1, INAC, 17 rue des Martyrs, 38054 Grenoble Cedex 9, France}
\affiliation{ESEME, PMMH-ESPCI, 10, rue Vauquelin, 75231 Paris Cedex 5, France}
\date{\today}

\begin{abstract}
We report quench cooling experiments performed
with liquid O$_2$ under different levels of gravity as provided by a magnetic
gravity compensation setup. A copper disk is quenched from 270K to 90K. It is found that when gravity is zero, the cooling time is abnormally long in comparison with any other gravity level. This phenomenon can be explained by the insulation effect of the gas surrounding the disk. The liquid subcooling is shown to drastically improve the heat exchange, thus reducing the cooling time (about 20 times). The effect of subcooling on the heat transfer is analyzed at different gravity levels. It is shown that such type of experiments cannot be used for the analysis of the critical heat flux (CHF) of the boiling crisis. The minimum heat flux (MHF) of boiling is analyzed instead.
\end{abstract}
\pacs{47.55.np, 68.03.Fg, 47.15.gm}
\maketitle

\section{\label{intro}Introduction}

The rocket engines for spatial applications need to be able to restart under
low gravity conditions to place satellites on different orbits. The solid
fuel engines are currently used for this purpose. The cryogenic fuel like
hydrogen and oxygen is however much more advantageous energetically and is
currently under study for next generations of the rocket engines. There
is however an important limitation. After some period of inactivity, the
temperature of the fuel injectors of such engines may rise above the
temperatures at which the fuel may be in the liquid state because of the
solar radiation. The engine cannot function before the injectors are not
chilled down. The injectors are usually cooled by putting them into the
contact with the cryogenic fluids (i.e. with the fuel), i.e. by the quench
cooling. One thus needs to master the cooling time as a function of
different parameters. The understanding of the latter phenomenon under
reduced gravity conditions thus becomes important. In the present article we
study the quenching of a hot metal piece in cryogenic fluids as a function
of several parameters like the gravity level.

The prediction of cooling time necessitates the knowledge of the heat
transfer coefficient which depends on temperature. To measure the influence
of the gravity level on the value of the heat transfer coefficient, we have performed experiments in a liquid oxygen-filled cell. A
copper disc was quickly immersed into this bath and cooled down (quenching
method). The Earth gravity was compensated by the magnetic force in the OLGA
(Oxygen Low Gravity Apparatus) facility located at CEA/Grenoble in France.

The objective of this article is three-fold. We aim to show that (i) one
cannot measure the critical heat flux of the boiling crisis in a quenching
experiment and (ii) the subcooling improves greatly the cooling efficiency in microgravity that is otherwise very poor. Finally,
we measure the dependence of the minimum flux of boiling (i. e. the threshold of the departure from film to nucleate boiling) as a function of
gravity.

\subsection{The quench cooling}\label{qcool}

\begin{figure}[h]
\centering
\includegraphics[width=0.8\columnwidth,clip]{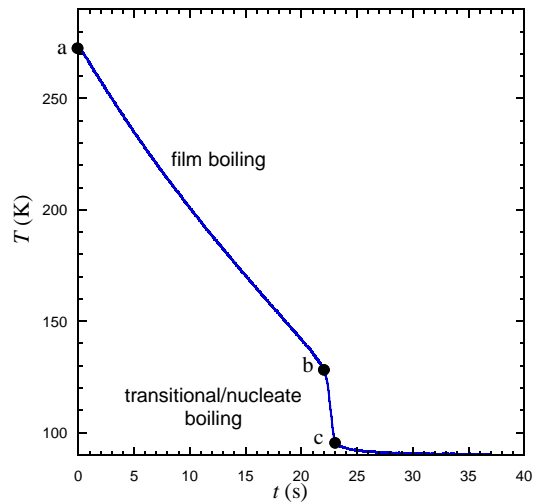}
\caption{Temperature evolution of a copper disk initially at about 270K dipped into O$_2$ bath at 90K under Earth gravity.} \label{CoolingEx}
\end{figure}

Consider what happens when a hot sample is dipped into a liquid. Fig.~\ref{CoolingEx}
presents a typical temperature evolution of a copper disk initially at about
270K dipped into a liquid O$_2$ bath at 90K under Earth gravity.

\begin{figure}[h]
\centering
\includegraphics[width=\columnwidth,clip]{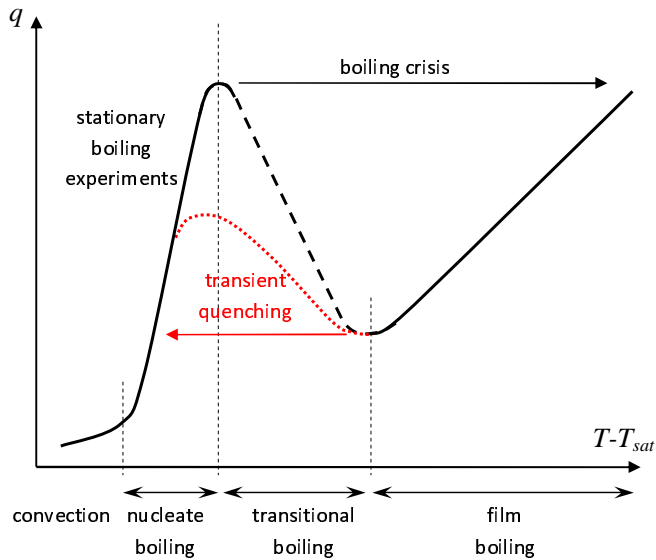}
\caption{(Color online) Schematic heat transfer curves for stationary boiling experiments and transient quenching.} \label{Nuk}
\end{figure}

At a first stage (a-b zone), the temperature decreases slowly because the
solid surface is completely covered by a layer of vapor which insulates the
solid from the liquid. This is the film boiling regime, during which the
vapor bubbles are generated at the vapor-liquid boundary. When the surface
temperature becomes low enough, the liquid may come into direct contact with
the sample and the bubbles start nucleating directly at the solid surface.
This signifies the beginning of the nucleate boiling regime which provides
much more efficient heat transfer and the temperature decreases very quickly
(b-c zone). At the final stage, the solid temperature is not any more
sufficiently high to induce boiling; the solid temperature approaches that
of the liquid O$_2$ bath due to convection. The evolution of heat transfer
regimes can be traced on the Nukiyama diagram (Fig.~\ref{Nuk}) presenting
the heat flux $q$ from the heater as a function of the sample temperature $T$.
The system evolves from the right (film boiling regime) to the left in Fig.~\ref{Nuk}. The moment where the direct solid-liquid contact begins to
occur corresponds to the minimum heat flux (MHF). From this moment, the
transition from film to nucleate boiling starts. The line along which the
system evolves depends on the thermal inertia of solid. If it was
vanishingly small, a transition to the nucleate boiling regime would occur
instantaneously (leftwise directed arrow in Fig.~\ref{Nuk}). In reality, a
finite time is necessary to cool down the solid and the system evolves along
the dotted curve. The heat flux attains a maximum value at a point where no
part of the solid is covered by the persisting vapor film. The nucleate boiling regime
occurs from that moment on. It is very important to distinguish the
quenching curve from the boiling (solid) curve.

The maximum heat flux (called critical heat flux, CHF) can only be recovered
in the stationary boiling experiments (see, e.g. \cite{PRL06}) performed at
fixed heat flux during each experiment. Till 1960s, the quenching experiments
were considered to be suitable for the CHF determination \cite{Pilling20,Merte64}. With
the development of the quenching studies for the metallurgy (steel
quenching), it became clear that only MHF may be obtained by this method
\cite{Kobasko97}, the maximum flux value depending strongly on the sample
thermal inertia. The actual CHF may only be obtained for the samples of the
large enough thermal inertia. However the CHF may hardly be studied in that
case, which is evident from the following consideration.

During the cooling, the heat $Q$ transferred from the sample to the liquid bath
during the time $\Delta t$ may be expressed as
\begin{equation}
\label{eq1}
Q = mC_P \left( T \right)\left( {T\left( t
\right)-T\left( {t+\Delta t} \right)} \right),
\end{equation}
where $m$ is the mass of the disk, $T$ its temperature, and $t$ is
the time. The specific heat of copper $C_P$ depends on $T$,
\[
\begin{split}
C_p ( T )= -1.355\times
10^{-7}T^4+ 1.303\times
10^{-4}T^3\\- 4.798\times
10^{-2}T^2+ 8.331T-
217.964
\end{split}
\]
The transferred heat may also be expressed with the heat transfer
coefficient $h$,
\begin{equation}
\label{eq2}
Q = hS\left( {T-T_L } \right)\Delta t,
\end{equation}
where $S$ is the total surface of the sample and $T_L$ is the temperature of the
liquid bath. By equating these expressions, one finds
\[
h=\frac{mC_P [T(t)][ T(t)-T(t+\Delta t) ]}{S\Delta t[T(t)-T_L]}
\]
and the heat flux $q=Q/(S\Delta t)$.

This method of measurements is valid only if the temperature of the sample
is uniform during the cooling. For this, the Biot number
\[
Bi=\frac{hL_c }{\lambda }
\]
must be lower than 0.1~\cite{Kreith67}. Here $L_c$ is the characteristic length (a half thickness of the disk); $\lambda$ is the copper thermal conductivity. It has been checked that the criterion $Bi<0.1$ holds for all described
experiments.

One can see now that the requirements of small $Bi<0.1$ and large thermal
inertia of the sample are contradictory. This explains why the quench
method is hardly suitable for the CHF study. The objective of this article is the analysis of the MHF rather than of CHF. Besides we study the cooling dynamics and the impact of subcooling on it at various gravity levels.

\section{Experimental apparatus}

\begin{figure}[h]
\centering
\includegraphics[width=\columnwidth,clip]{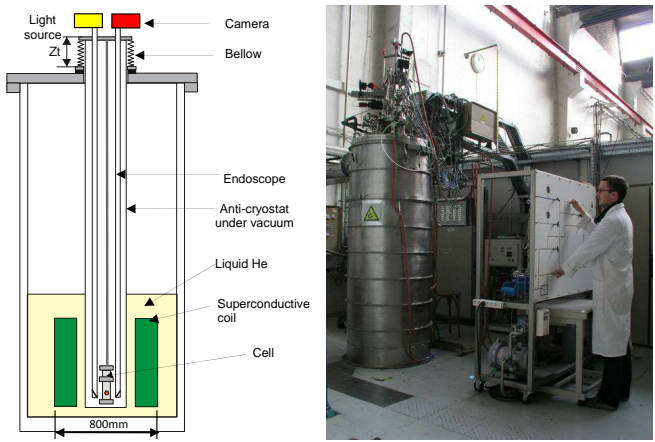}
\caption{(Color online) A sketch and a general view of the OLGA facility.} \label{OLGA}
\end{figure}

Experiments were performed in the OLGA facility (Fig.\ref{OLGA}). It is equipped with a
superconductive coil and creates a magnetic force strong enough to compensate the buoyancy force in O$_2$.

\begin{figure}[h]
\centering
\includegraphics[width=\columnwidth,clip]{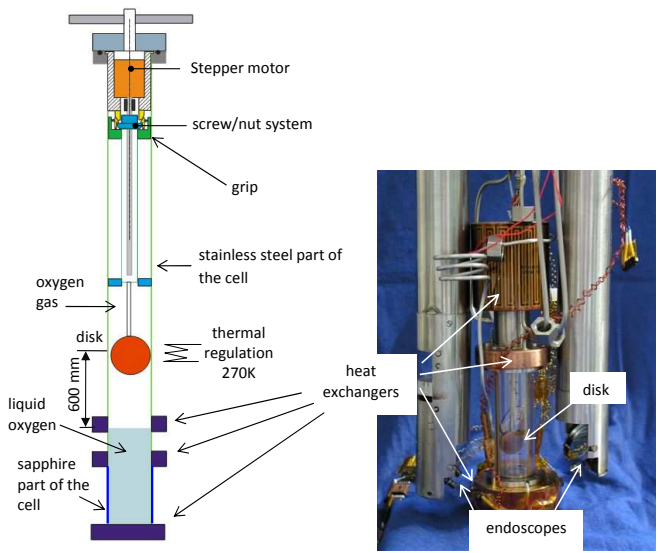}
\caption{(Color online) A sketch of the cell and a photo of its lower part (the upper part of the cell is removed). The disk is shown in the upper position on the sketch and in the lower on the photo.} \label{Cell}
\end{figure}

Oxygen is contained in the experimental cell (Fig.\ref{Cell}) placed
in an anti-cryostat inserted into the 300~mm bore of the
superconductive coil. Two endoscopes are used for observation: one for the light and another for the camera.

The lower part of the experimental cell is made of a sapphire tube of 30~mm diameter and 100~mm
length closed by two copper flanges equipped with heat exchangers, inside which
the gaseous cold helium may flow. The temperature of the two flanges is controlled
within 0.01K with heaters. The upper part of the cell is the stainless steel tube. The pressure in the cell is measured
with an accuracy of 10~mb. Liquid oxygen is obtained by condensing pure
O$_2$ gas (99.995{\%}) in the cell. The cell is brought to the desired temperature
by regulating the power of heaters and helium flow rate in the heat exchangers
that are situated at the top and at the bottom of the transparent part of
the cell. 

The disk is made of pure copper. Its diameter is 20.02~mm and its
thickness is 2.98~mm. The weight of the disk is 7.68~g, its surface in contact
with liquid is $S=8.07$~cm$^2$ and its RMS roughness is
0.8~$\mu$m. A hole was drilled to introduce a platinum thermometer
(supplier: Heinz Messtechnic, model W60-50, class A; diameter 1.8~mm, length
11~mm) in its heart. After each quench experiment, the disk may be risen into a hot zone that situates 0.6~m above the
transparent part of the cell with a screw/nut system driven by a stepper
motor. The hot zone of the cell is regulated at 270~K. Once the disk attains this temperature,
the grip is loosened again and the disk drops into the liquid. Meanwhile, the disk temperature is acquired at a frequency of 100~Hz.

\section{Gravity compensation}
\label{sec:gravity}

The principle of magnetic compensation of gravity in a pure non-magnetic
substance is an application of the magnetic volume force opposite to the
Earth gravity \cite{B&T91,Cryo02,MST11}.

This force per unit volume is given by the expression
\begin{equation}\label{magn}
\vec {F}_m =\frac{\chi _m }{2\mu _0 }\nabla(B^2),
\end{equation}
where $B$ is the magnetic field, $\mu _0$ is the magnetic permeability of
vacuum and $\chi _m$ the magnetic susceptibility of the material.

Since $\chi _m $ is proportional to the density $\rho$ of the nonmagnetic
substance, one may introduce a quantity $\alpha =\chi _m /\rho$ independent
of the density. It means that the buoyancy force is compensated if one
applies the field with $dB^2/dz=2\mu _0 g/\alpha$, where
$g$ is the Earth gravity acceleration and $z$ is the vertical coordinate. For O$_2$ at
90K, $2\mu _0 g/\alpha\sim 8$~T$^{2}$/m. These
conditions are achieved in OLGA facility near the bottom of the
superconductive coil.

\begin{figure}[h]
\centering
\includegraphics[width=5cm,clip]{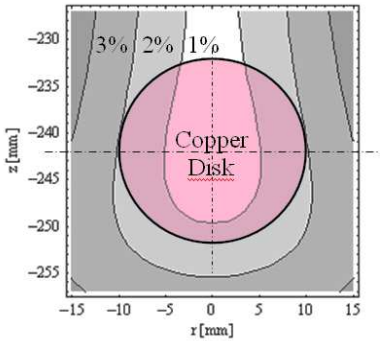}
\caption{(Color online) Iso lines of the residual acceleration modulus in {\%} of $g$ inside the cell of OLGA at the ending point of the disk fall at $0g$.} \label{Field}
\end{figure}

It was demonstrated \cite{Quettier} that the magnetic force (\ref{magn}) cannot be
made constant in a volume. In our experimental cell, Earth gravity may be exactly
compensated at a single point called complete compensation point. Such conditions are called hereafter $0g$ and the corresponding coil current is denoted $I_{0g}$. In the vicinity of this point, a residual
acceleration acts on O$_2$ \cite{MST09}. As the magnetic force is a way too small to
compensate the gravity for copper, the disk falls with the Earth gravity
acceleration. Its stop point situates at $z=-242$~mm with respect to the
center of the coil. At this location, Fig.~\ref{Field} gives the calculated iso-values of residual gravity
acceleration modulus at $0g$ conditions. One can see that at the edge of
the disk, the residual acceleration is smaller than 2\% of $g$.

\begin{figure}[h]
\centering
\includegraphics[width=4cm,clip]{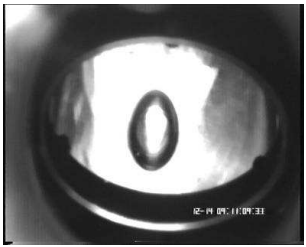}
\caption{Image of O$_2$ bubble at $0g$ inside the cell at equilibrium as seen in the periscope mirror used to adjust the position of the exact compensation point with respect to the cell. For this case, the temperature is 92.24K and the current in the coil is $I_{0g} =241$~A. The elongated shape of the bubble is due to the weak coupling of the magnetic field with the gas-liquid interface shape \cite{MST11}.} \label{b0g}
\end{figure}

To achieve $0g$ at
the required place inside the cell, a single bubble is generated by the boiling process with a heater
situated at the bottom of the cell. Then the current in the coil is adjusted
until the bubble center coincides with the optimal previously calculated
point (Fig.~\ref{b0g}).

The effective gravity level \cite{MST11} may be varied by changing the coil current $I$. It is known \cite{WunenH2} that the effective gravity level $g^*$ at the geometrical point of complete compensation is given by the equation
\[
g^* =\left( {1-\frac{I^2}{I_{0g} ^2}} \right)g.
\]

\begin{figure}[h]
\centering
\includegraphics[width=\columnwidth,clip]{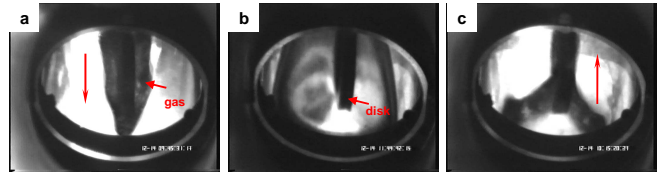}
\caption{(Color online) Film boiling at different gravity (its direction is shown with vertical arrows) at initial liquid O$_2$ temperature (before disk drop) 92.24K. a) $I=241$~A: the residual gravity is directed downwards and the gas surrounding the disk slowly rises. b) $I=245.18$~A: the gas recondenses without rising and surrounds the disk during all the cooling time (zero gravity). c) $I=252.7$~A: the residual gravity is directed upwards and the gas goes downwards.} \label{FilmBoil}
\end{figure}

The magnetic susceptibility of oxygen varies with temperature, and $I_{0g}$ depends weakly on the liquid temperature. For the isothermal conditions, the currents for the complete gravity compensation are $I_{0g}=239.7$~A at 90.07K
(pressure of 1~bar, at saturation) and $I_{0g}=241$~A at 92.24K (subcooling 5K for
the pressure of 2~bars). When the disk is dropped into liquid O$_2$, the liquid is slightly heated up, resulting in a small decrease of the magnetic susceptibility. By observing the movement of the gas generated by boiling
around the disk (Fig.~\ref{FilmBoil}), we could see that the current had to be slightly
increased to obtain $0 g$ in the cell during the disk cooling. In this case,
$I_{0g}=242.1$~A at 90.07K, and $I_{0g}=245.18$~A at 92.24K.

\section{Experimental results}

\begin{figure}[h]
\centering
\includegraphics[width=0.8\columnwidth,clip]{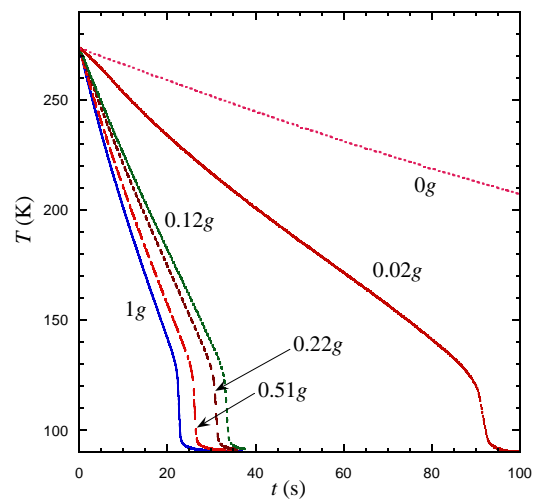}
\caption{(Color online) Evolution of the copper disk temperature dipped into O$_2$ bath at different gravity levels for the saturated case.} \label{FTsat}
\end{figure}

Quenching experiments were performed for two O$_2$ thermodynamic
conditions: saturated O$_2$ and subcooled O$_2$ with respect to its
saturation temperature for given pressure that was controlled independently
since the cell was connected to a large oxygen-filled vessel.

\subsection{Case of saturated liquid}

For saturated O$_2$ experiments, the temperature of the cell was regulated at
90.07K and the vessel pressure was maintained at 1 bar (which is the
saturation pressure corresponding to this temperature). 15 quenches were
performed.

\subsubsection{Cooling curves of the disk}

Fig.~\ref{FTsat} shows the cooling curves of the disk for different levels of gravity.
We can see that the lower the gravity, the larger is the cooling time.
The $0g$ cooling time is very large with respect to other cases and the corresponding curve is truncated: it took more than 10 minutes to cool down the disk. The complete curve is shown in the inset to Fig.~\ref{T0g} below.

\subsubsection{Heat transfer curves}

\begin{figure}[h]
\centering
\includegraphics[width=0.8\columnwidth,clip]{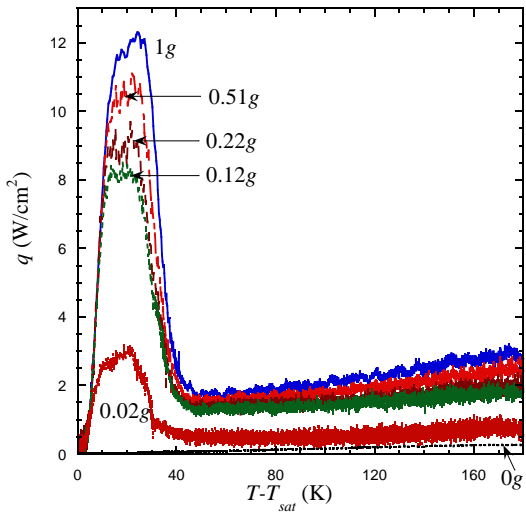}
\caption{(Color online) Heat transfer curves for saturated O$_2$ quench cooling at 1 bar and different gravity levels.} \label{NukSat}
\end{figure}

Using the above described method, we have calculated the heat
transfer curves for different levels of gravity (Fig.~\ref{NukSat}). The film boiling
points were calculated by making a centered moving average on 50 points for
the film boiling and from 10 to 20 points for the transitional/nucleate
boiling. One can clearly see the influence of the gravity level on the transferred
heat flux in film boiling regime and on the maximum and minimum heat flux values.

\begin{figure}[h]
\centering
\includegraphics[width=0.8\columnwidth,clip]{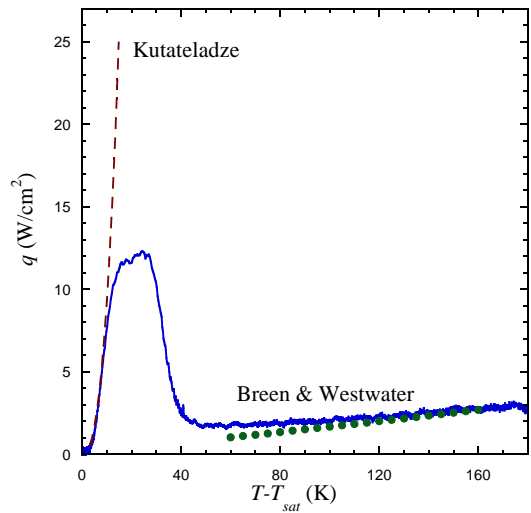}
\caption{(Color online) Comparison of the Kutateladze (broken line) and Breen{\&}Westwater (circles) correlations with our data from Fig.~\ref{NukSat} corresponding to $1g$ (solid line).} \label{Comp}
\end{figure}

Fig.~\ref{Comp} is a comparison of Kutateladze and Breen {\&} Westwater correlations \cite{NBS65}
with our experiment under $1g$.
One can see that the quench method gives the maximal heat flux $\sim 12$~W/cm$^{2})$, which is two times lower
than the CHF value 24.9~W/cm$^{2}$ obtained with the Kutateladze correlation. The
latter value reasonably agrees with the experimental value (22-23~W/cm$^{2}$) obtained in pool boiling experiments \cite{Lyon65}. The low value of the maximum heat flux reflects the transient nature of the quench cooling as discussed in Fig.~\ref{Nuk}(sec. \ref{qcool}).

\subsection{Case of subcooled liquid}
\begin{figure}[h]
\centering
\includegraphics[width=4cm,clip]{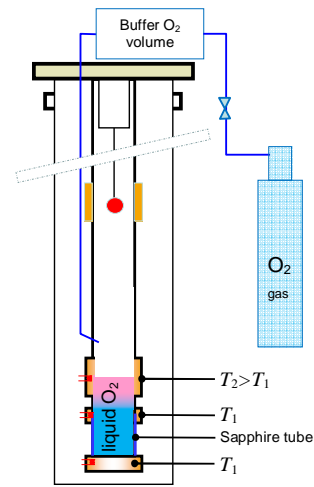}
\caption{(Color online) Sketch showing the principle used for subcooling experiments.} \label{SubCool}
\end{figure}

To measure the subcooling influence, experiments where performed for several
subcooling values. The subcooling was obtained with the following
method (cf. Fig.~\ref{SubCool}). The temperatures of two lower heat exchangers of the cell (cf. Fig.~\ref{Cell})
are regulated at the same value $T_1$. The temperature $T_2$ of the upper heat exchanger
is regulated at a higher value equal to the saturation temperature corresponding to the
gas pressure. Under these conditions, the equilibrium level of liquid forms in the region of cell maintained at the temperature $T_2$. A thermal gradient exists in the liquid between two upper heat exchangers. The liquid is isothermal at $T_1$ in between two lower heat exchanges. The final drop point of disk is situated in this initially isothermal region.

To achieve the 5K subcooling, $T_1=92.24$K and $T_2=97.24$K. The latter temperature corresponds to the
pressure of 2 bars. For the 7K subcooling (see below), $T_1=90.24$K.

\subsubsection{Cooling curves}

\begin{figure}[h]
\centering
\includegraphics[width=0.8\columnwidth,clip]{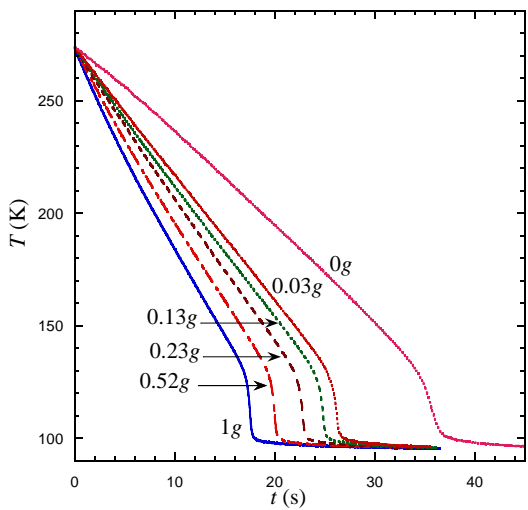}
\caption{(Color online) Evolution of the disk temperature at different gravity levels for the 5K subcooled case.} \label{T5K}
\end{figure}

Fig.~\ref{T5K} shows the raw cooling curves of the disk for 6 levels
of gravity. The temperature of cell
was regulated at 92.24K and the pressure was adjusted at 2~bars which results in a
subcooling of 5K. The temperature acquisition frequency was 100~Hz. Similarly to the saturated case, the lower the gravity level, the
longer is the cooling time. Unlike the saturated case, the $0g$ cooling time is only twice larger than that of the $1g$ case.

\subsubsection{Impact of subcooling}

\begin{figure}[h]
\centering
\includegraphics[width=0.8\columnwidth,clip]{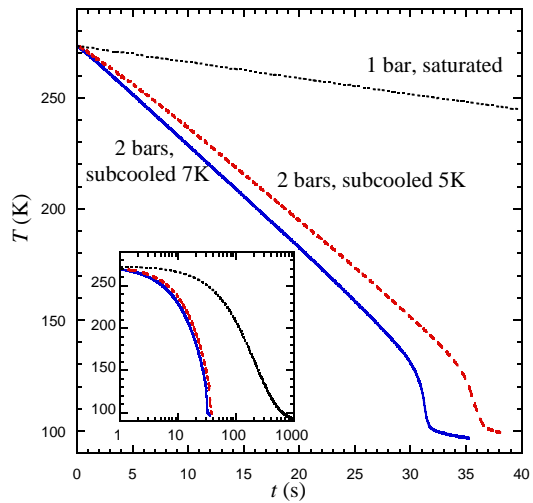}
\caption{(Color online) Evolution of the disk temperature at different subcooling levels at $0g$. The same curves are shown in the inset in the semi logarithmic scale.} \label{T0g}
\end{figure}

Fig.~\ref{T0g} shows that the subcooling reduces drastically (20 times) the general cooling
time at $0g$ while the effect at any nonzero gravity is much weaker (cf. Figs.~\ref{FTsat} and \ref{T5K}). Such a reduction may be explained as follows. Fig.~\ref{FilmBoil}b shows the formation at $0g$ of a vapor bubble that surrounds the solid and completely insulates it from the liquid. The bubble can recondense if the liquid is subcooled and cannot recondense if the liquid is saturated. For the saturated case, the heat exchange is thus produced via the weak heat conduction through the vapor and through the almost non-conductive at cryogenic temperatures stainless steel rod that supports the disk (cf. Fig. \ref{Cell}). When the liquid is subcooled, the recondensation of the vapor at the external bubble interface is very efficient and the gas layer surrounding the disk is much thinner. This effect improves the heat exchange thus favoring cooling.

\subsubsection{Heat transfer curves}

\begin{figure}[h]
\centering
\includegraphics[width=0.8\columnwidth,clip]{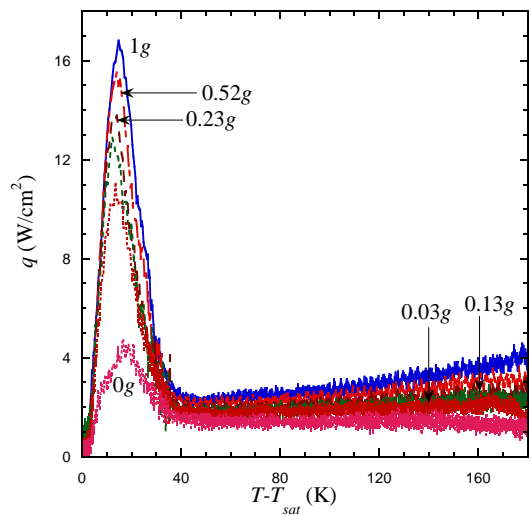}
\caption{(Color online) Heat transfer curves for the 5K subcooling case at different gravity levels.} \label{Nuk5K}
\end{figure}

We have plotted the heat transfer curves for 6 levels of gravity (Fig.~\ref{Nuk5K}) using
the same method as the one used for the saturated case. One can see that unlike the saturated quench,
the maximum and the minimum heat flux values are both nonzero for the $0g$ case.

\subsection{Minimum heat flux}
\begin{figure}[h]
\centering
\includegraphics[width=6cm,clip]{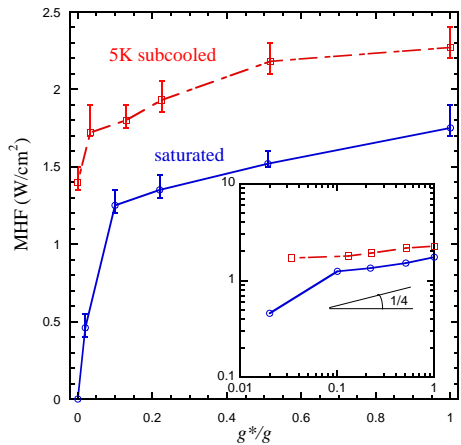}
\caption{(Color online) Minimum heat flux versus gravity level for saturated and subcooled O$_2$. Inset: same in log-log scale; the 1/4 slope is shown for comparison.} \label{MHF}
\end{figure}

Fig.~\ref{MHF} shows the minimum heat flux as a function of the gravity level. The
minimum value was obtained by fitting the raw data with a second degree polynomial. The MHF occurrence is usually associated with the Taylor instability of the vapor film \cite{Lienhard80,Ramilison87} so that $q_{MHF}\sim g^{1/4}$ should be satisfied. Fig.~\ref{MHF} shows an exponent close to this value.

\section{Conclusion}

Quench cooling has been analyzed under different gravity levels. Earth gravity was compensated with magnetic forces. Measurements were performed
by quenching a copper disk of 20 mm diameter and 3 mm thickness from 270K to 90K. The analysis of different boiling regimes has shown that the study of the critical heat flux (CHF) via quenching experiments is hardly possible because of the transient nature of the quench cooling: the maximum heat flux value is twice smaller than the CHF measured by other authors in the stationary boiling experiments.

It has been shown that cooling via quenching in microgravity takes usually exceedingly long time because the vapor generated during the boiling envelopes completely the cooled body (film boiling regime), thus thermally insulating it from the liquid. It is shown that the artificial subcooling (for instance, by transient pressurization) can drastically accelerate the cooling. This speeding up is explained by the recondensation of the vapor envelope at the vapor-liquid interface in case of subcooling.

The dependence of the heat transfer and, in particular, of the minimum heat flux of boiling (MHF) on gravity is analyzed. As expected, the heat transfer improves with the gravity level. The MHF gravity dependence
shows an exponent close to that of the Zuber theory (exponent 1/4).

\begin{acknowledgments}
The authors gratefully acknowledge Jer\^ome Chartier, Jean-Marc Mathonnet, Patrick Bonnay and
Stephane Garcia (CEA/SBT) for their contribution throughout this work.

This work was partially supported by CNES.
\end{acknowledgments}

%

\end{document}